\documentclass{aa}

\usepackage{graphicx}
\usepackage{txfonts}
\usepackage{natbib}
\usepackage{color}

\newcommand{\Msun}{\mbox{\,M$_\odot$}}
\newcommand{\Rsun}{\mbox{\,R$_\odot$}}
\newcommand{\Lsun}{\mbox{\,L$_\odot$}}
\newcommand{\vunit}{\mbox{\,km\,s$^{-1}$}}
\newcommand{\mic}{\mbox{$\,\mu$m}}
\newcommand{\pion}[2]{{#1}\,{\sc {#2}}}

\newcommand{\Tef}{\mbox{$T_{\rm e}$}}
\newcommand{\logg}{\mbox{$\log g$}}
\newcommand{\ltsimeq}{\raisebox{-0.6ex}{$\,\stackrel
        {\raisebox{-.2ex}{$\textstyle <$}}{\sim}\,$}}

\newcommand{\kic}{\mbox{KIC\,9832227}}

\begin{document}

   \title{Infrared spectroscopy of the merger candidate KIC\,9832227}

   \author{Ya. V. Pavlenko{\thanks{E-mail:yp@mao.kiev.ua}}
          \inst{1}
          \and
          A. Evans\inst{2}
          \and
          D. P. K. Banerjee\inst{3}
          \and
          J. Southworth\inst{2}
          \and
          M. Shahbandeh\inst{4}
          \and
          S. Davis\inst{4}}

 \institute{
 {Main Astronomical Observatory, Academy of Sciences of the Ukraine, Golosiiv Woods, Kyiv-127, 03680 Ukraine}
 \and
 {Astrophysics Group, Keele University, Keele, Staffordshire, ST5 5BG, UK }
 \and
 {Physical Research Laboratory, Navrangpura, Ahmedabad, Gujarat 380009, India}
         \and
         {Department of Physics, Florida State University, 77 Chieftain Way, Tallahassee, FL 32306-4350, USA} }

   \date{Received; accepted}


  \abstract
   {It has been predicted that the object \kic\ -- a contact binary star -- will undergo a merger in
   $2022.2\pm0.7$. We describe the near infrared spectrum of this object as an impetus to obtain pre-merger data.}
   {We aim to characterise (i)~the nature of the individual components of the binary and 
   (ii)~the likely circumbinary environment, so that the merger -- if
   and when it occurs -- can be interpreted in an informed manner.}
   {We use infrared spectroscopy in the wavelength range 0.7\mic--2.5\mic, to which we fit model atmospheres to represent
   the individual stars. We use the binary ephemeris to determine the orbital phase at the time of observation.}
   {We find that the infrared spectrum is best fitted by a single component having effective temperature
   5\,920~K, $\log{[g]}=4.1$ and solar metallicity, consistent with the fact that the system was observed at conjunction.}
   {The strength of the infrared H lines is consistent with a high value of $\log{g}$, and the strength of the 
   \pion{Ca}{ii} triplet indicates the presence of a chromosphere, as might be expected from rapid
   stellar rotation. The \pion{He}{i} absorption we observe likely arises in He excited by
   coronal activity in a circumstellar envelope, suggesting that the weakness of the \pion{Ca}{ii} triplet is 
   also likely chromospheric
   in origin.}


   \maketitle
%

\section{Introduction}

There has recently come to light a new class of eruptive events
whose outbursts can be attributed to a stellar  merger.
These events generally display high luminosity at maximum \citep{bond03}
($M_{\rm bol} \sim -10, M_v \ltsimeq -9$) as 
evidenced by the occurrence of LRVs in M31 \citep{bond03,
will15} and in other galaxies \citep{kasl12,smith16}.
While the nature of these eruptions was initially unclear
(with nova eruptions, planet-swallowing stars, very late thermal pulses having been suggested)
the ``best-bet'' scenario -- based on the behaviour of V1309~Sco -- is {\em the merger of two stars}.

V1309~Sco is the ``Rosetta Stone'' of stellar mergers, in that its pre-eruptive
behaviour and the subsequent eruption are not only consistent with, but conclusively point to,
a stellar merger event in a contact binary. The progenitor had an orbital
period of 1.4~day that decreased up to the 2008 eruption \citep{tyle11}.
The eruptive event that marked the merger began in 2008 March,
reaching peak luminosity $\sim3\times10^4$\Lsun\ \citep{tyle11}.
Infrared (IR) observations \citep{mcco14} show that it had an IR excess 
at least a year before the merger. \cite{pejc14} has shown
 that V1309 Sco experienced mass loss through the outer
Lagrangian point, which eventually obscured the binary.
\cite{zhu13} note that the presence of a
significant amount of dust around V1309~Sco suggests that the ejecta
following a stellar merger is an ideal environment for dust formation and growth; they
showed that $\sim5.2\times10^{-4}$\Msun\ of silicate and iron dust
was produced in the merger.

\kic\ was identified by \cite{moln15} as a potential merger
candidate, and we present here a near IR spectrum of this candidate merger object.

\section{\kic}
\kic\ is both an eclipsing and a contact binary, with an orbital period 0.458~days and amplitude $\sim0.2$~mag in the optical.
\citeauthor{moln17} (2017, hereafter M17) found that the orbital period as determined from {\it Kepler} data was
significantly shorter than in previous years, and that the period derivative was rapidly decreasing, a sure 
signature of an impending merger; they estimate the time of merger to be $2022.2\pm0.7$.
M17 determine that the primary (secondary) star has mass 1.4\Msun\ (0.32\Msun), radius 2.6\Rsun\ 
(0.78\Rsun), $T_{\rm eff} = 5\,800$~K (5\,920~K) and $\log{g} = 4.19$ (4.10).
There is also some evidence for the presence of a third component in the system, with orbital period $590\pm8$~days and
$M\sin{i}=0,11$\Msun. A detailed discussion of the merger potential of \kic\ is given by M17.


\section{Observations}

\subsection{IRTF observation}

\begin{figure*}
   \centering
   \includegraphics[width=7.5cm]{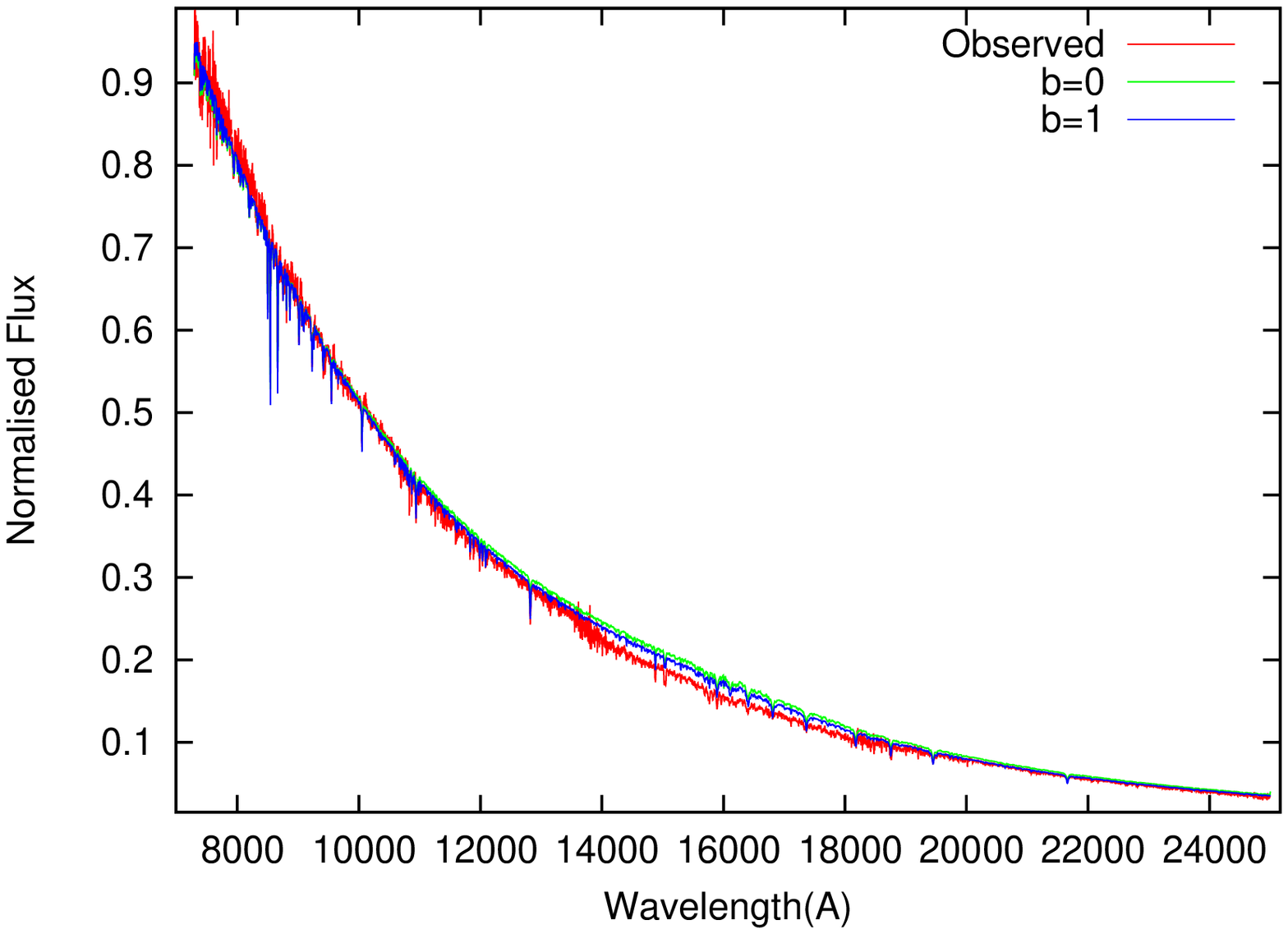}
    \includegraphics[width=7.5cm]{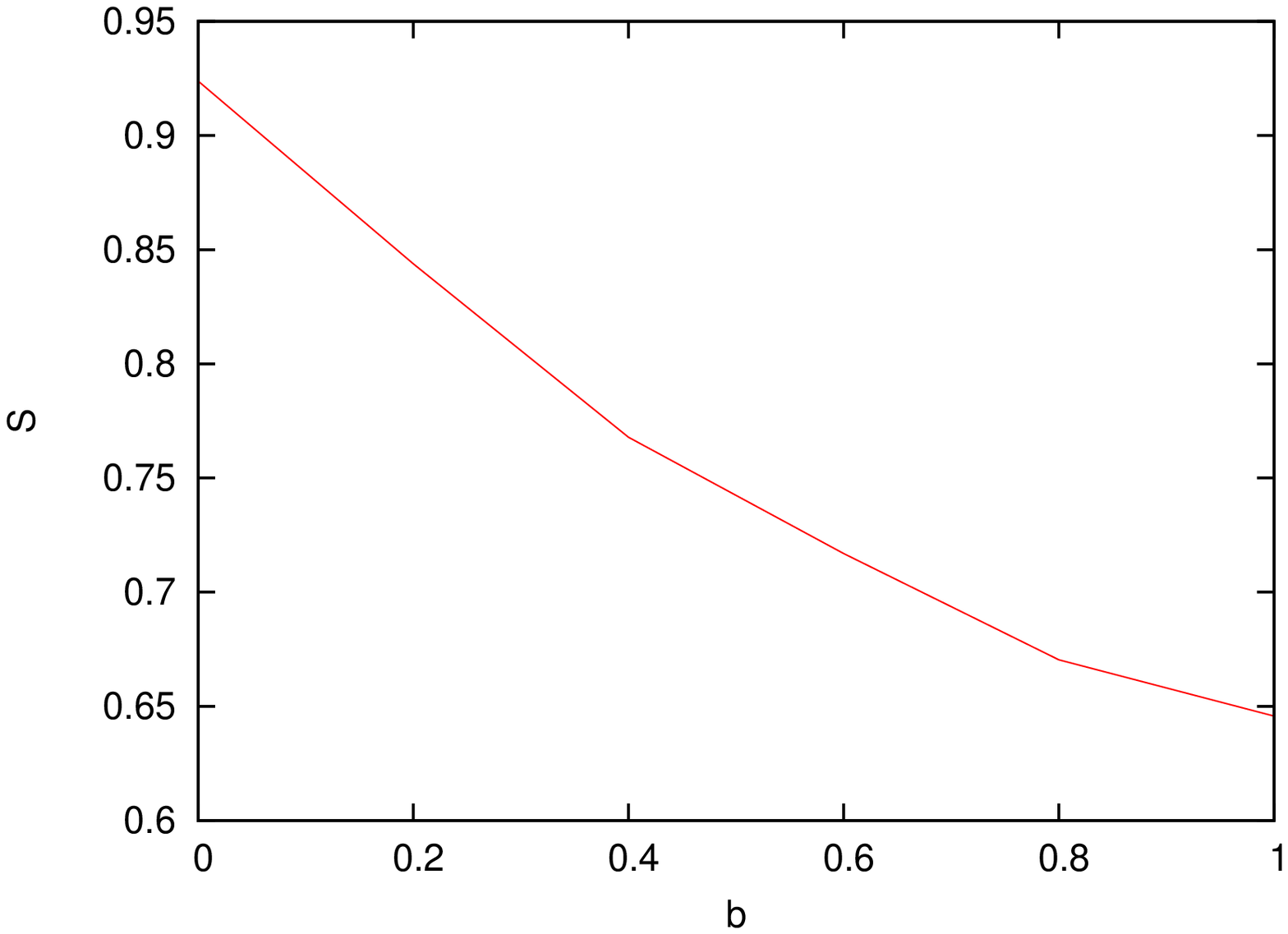}
   \caption{Left: dereddened IRTF spectrum of \kic, together with the fit with $b =1$; the case  
   $b =0$ is also shown for comparison. See text for details.
    Right: minimisation parameter $S$ as the function of $b$.}
   \label{_S}
    \end{figure*}

\kic\ was observed with the SpeX spectrograph \citep{rayner}
on the 3\,m NASA Infra-Red Telescope
Facility (IRTF), Hawaii, on 2017 July 7.54 UT (MJD 57941.54). SpeX
was used in the cross-dispersed mode using a $0.5''\times15''$ slit
resulting in a spectral coverage  $0.77-2.50$\mic, 
at resolving power $R=\lambda/\delta\lambda=1\,200$. The total integration time
was 717 seconds. The A0V star HD194354 was used as the
telluric standard. The data were reduced and calibrated using the Spextool software 
\citep{cushing}, and corrections for telluric absorption were performed using the IDL
tool Xtellcor (Vacca et al. 2003). 
The observed spectrum, which was dereddened by $E(B-V)=0.03$ (M17), is shown in Fig.~\ref{_S}.

\subsection{Complementary observations}

\begin{figure}
   \centering
   \includegraphics[width=7cm]{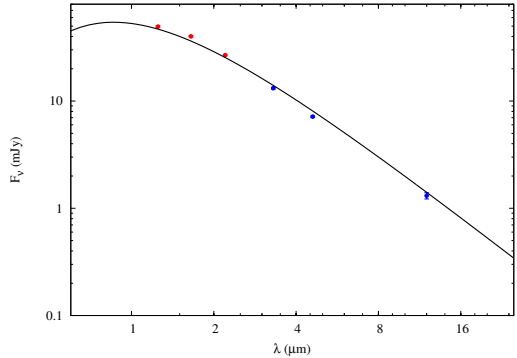}
   \caption{Infrared spectrum spectral energy distribution of \kic\ from
   2MASS (red) and WISE (blue) surveys.
   Black curve is a 5\,930~K blackbody, corresponding to the stellar components of \kic.
   \label{KIC_IRTF_WISE}}%
    \end{figure}

\kic\ is included in the 2{\it MASS} \citep{skru06} and {\it WISE} \citep{wrig10} IR surveys;
the source was not detected in other surveys, such as {\it IRAS}, {\it AKARI} and {\it Herschel} PACS.
The 2MASS and WISE fluxes are given in Table~\ref{IR} and shown in Fig.~\ref{KIC_IRTF_WISE};
\kic\ was not detected in {\it WISE} Band~4 (22\mic). 
The IR spectral energy distribution (SED) is photospheric out to at least 12\mic, indicating that dust -- 
if present in significant quantity as in V1309~Sco
-- has temperature $\ltsimeq425$~K.

\begin{table}
\caption{Infrared fluxes from 2MASS and WISE.}\label{IR}
\centering
\begin{tabular}{cccc}
\hline\hline
Survey & Band     & $\lambda$ ($\mu$m) & Flux (mJy) \\ \hline
 2MASS  &  $J$     & 1.25   &  $48.23\pm0.92$ \\
       &  $H$     & 1.65 &  $39.55\pm0.62$ \\
       &  $K_s$   & 2.2  &  $26.59\pm0.39$ \\
WISE   &  W1      & 3.3  & $13.13\pm0.28$ \\ 
       &  W2      & 4.6  & $7.14\pm0.14$  \\
       &  W3      &  12  & $1.31\pm0.09$ \\ \hline
\end{tabular}
\end{table}

\subsection{Orbital phase of \kic\ at the time of the IRTF observation}

\kic\ is an eclipsing binary: in order to interpret the IR spectrum we need to know the 
orbital phase and stellar velocity separation at the time of the IRTF observation. 
M17 do not give the full information necessary to deduce these quantities, so this process was non-trivial. 
We neglected the effect of the third body, which affects the orbital timings by only $\pm$0.002 cycles 
(Figure~10 in M17), and the proposed three-body solutions suggested by M17 because this body 
was not detected in the spectrum.

We adopted the parameters of the exponential fit given in Figure~12 of M17 and the linear orbital 
ephemeris from the Kepler Eclipsing Binary  
Catalogue\footnote{\texttt{http://keplerebs.villanova.edu/overview/?k=9832227}} \citep[see][]{kirk}. 
We further imposed a normalisation which forces the first point in the year 2004 to be at zero $O-C$. 
The resulting orbital phases correctly reproduce the $O-C$, instantaneous orbital period, 
and $dP/dt$ as plotted in Figures~12, 13 and 14 in M17.

The UTC time of mid-observation of our spectrum (2457942.04199) was converted into the 
BJD(TDB) timescale using the IDL routines from Eastman et al. (2010), giving a time of 2457942.04135. 
This corresponds to an orbital phase of 0.010, so our spectrum was obtained around a time of inferior 
conjunction. The velocity separation of the two stars at this time was negligible compared to their 
spectral line broadening, so the lines of the two stars are superposed in the spectrum.

\begin{figure*}
   \centering
   \includegraphics[width=7cm]{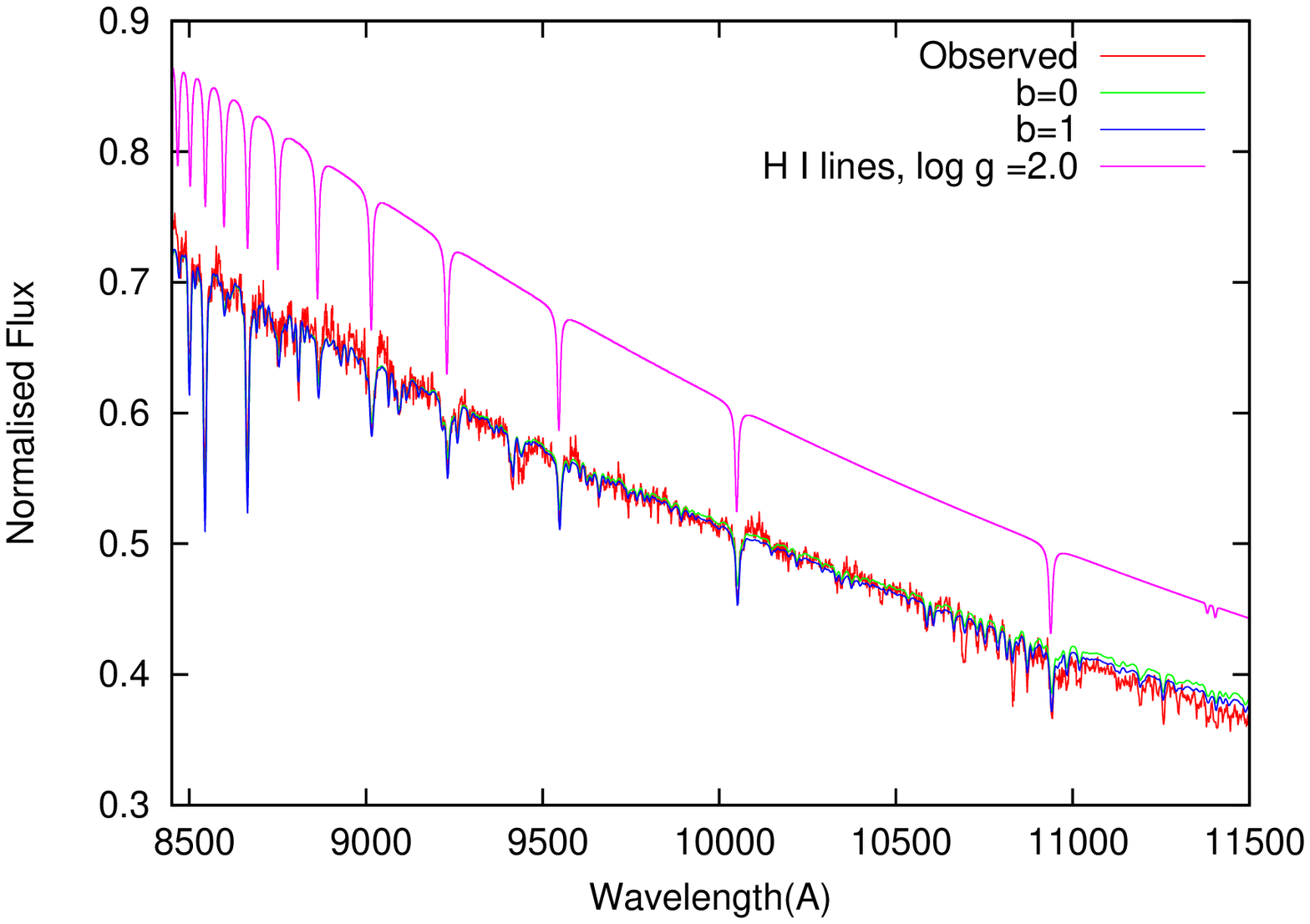}
      \includegraphics[width=7cm]{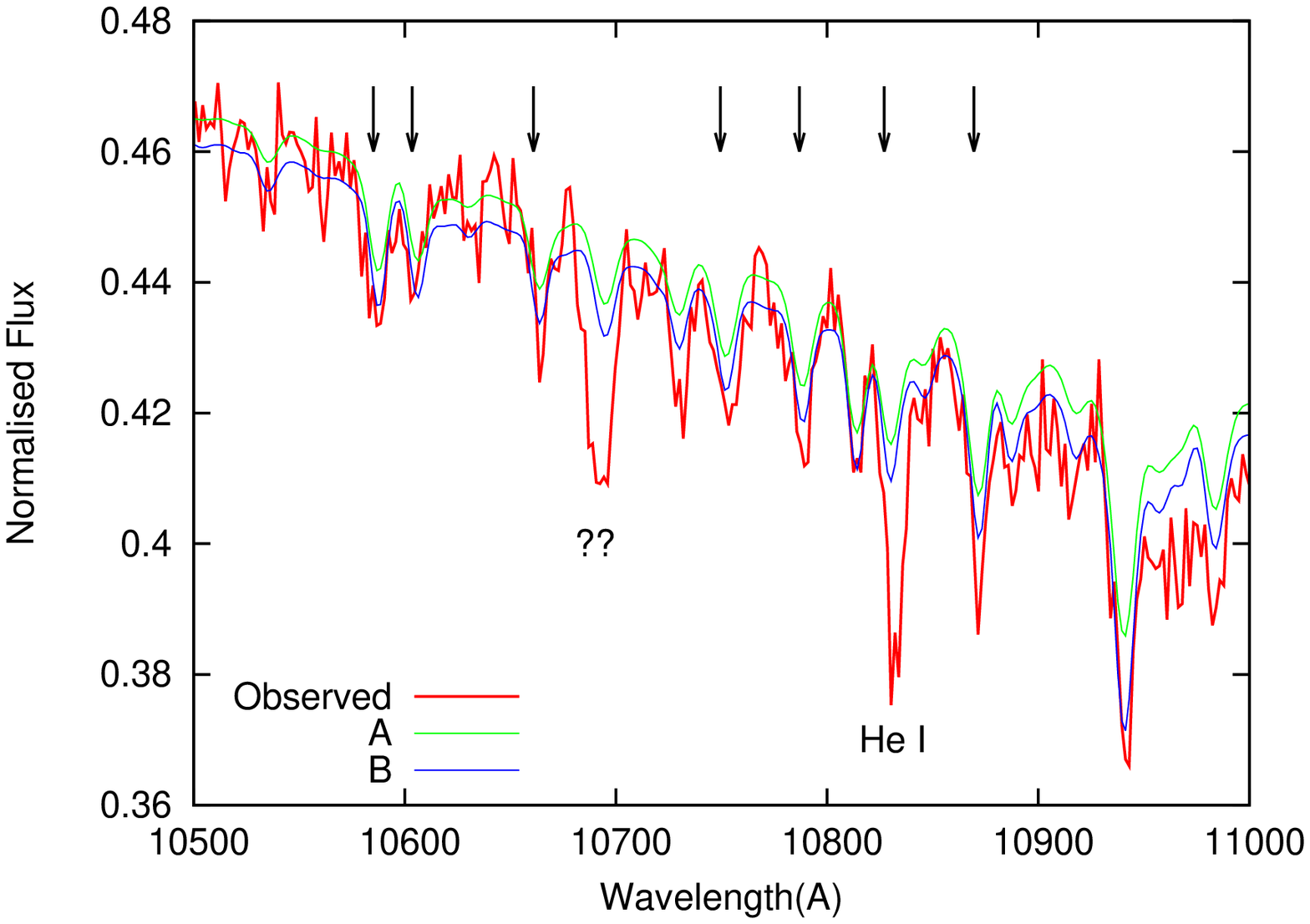}
   \caption{Left: best the fit to the observed 
   spectrum with $b =1$ in the region of the \pion{Ca}{ii} triplet and \pion{He}{i}; the case of $b=0$ is also shown,
   as is the hydrogen line spectrum computed for a model atmosphere with $\log{g} = 2.0$. 
   Right: fit to the spectrum containing Si I lines, marked by vertical arrows;
   the \pion{He}{i} line is also shown. Note the broad
   feature at 10\,700\,\AA, marked by '??'; see text.}
   \label{_fits}
    \end{figure*}
    
\section{Theoretical spectra and SEDs}

Synthetic spectra and SEDs
were computed using the WITA6 program for the SAM12 model 1D atmospheres  \citep{pavl03}
with \Tef(in K)/\logg/[Fe/H]= 5\,920/4.10/0.0 and 5\,800/4.19/0.0 for the hotter ($A$) 
and cooler ($B$) components, respectively, as determined by M17.
Significant molecular features are not seen in the observed spectrum, therefore 
we accounted for only atomic lines from the VALD3 database \citep{ryab15}. 
Our procedure for producing the theoretical 
spectra is described elsewhere \citep[see][]{pavl03, pavl08}. The
synthetic spectra were convolved with a gaussian profile having the instrumental 
broadening corresponding to the spectral resolution $R$,
and rotational profiles corresponding to $v\sin{i} =85$\vunit\ ($A$) and 150\vunit\ ($B$; M17). 
To account for rotational broadening of the spectra we used formulae given by \cite{gray76}.

\section{Results}
\label{results}
\subsection{The overall spectrum}

We computed theoretical spectral energy distributions for the $A$ and $B$ components, and created a grid of
combined spectra $F^t = b\times{F_A} +(1-b)\times{F_B}$, where $0<b<1$. We computed the best fits to the dereddened spectrum
for the cases $b=0.0, 0.2, 0.4, 0.6, 0.8, 1.0$; the procedure to determine the best fit used a $\chi^2$ minimising algorithm,
details of which are given elsewhere \citep[see][]{pavl08}. The minimisation parameter $S =
\sum(F_{\lambda}^{\rm t}-F_{\lambda}^{\rm obs})^2$ indicates a minimum $S$ at $b = 1$ (see Fig.~\ref{_S}).
This is consistent with the fact that the observed spectrum was obtained at conjuction, 
when the hotter component dominates the combined spectrum
(although a more compelling reason for not seeing the second component is the very similar
$T_{\rm eff}$). Furthermore, as the eclipses are very shallow, 
component $A$ dominates the spectrum at all orbital phases.

\subsection{Absorption lines in \kic's spectrum}

Generally speaking, both the intermediate resolution of the observed spectra ($R=1200$) and the rapid rotation of both components
complicate the use of atomic lines for abundance analysis. However, we see that practically all notable lines
in the near IR spectrum are well fitted with close-to-solar abundances; the likely errors do not exceed
0.3 dex. In the right panel of Fig.~\ref{_fits} we show a few well-fitted \pion{Si}{i} lines in the spectral range 10\,500-11\,000\,\AA. 
We find fits of the same quality to metallic lines in other spectral regions. 

We draw particular attention to the modelling of the moderately strong hydrogen lines, to which our model 
provides a satisfactory fit. In the left panel of Fig. \ref{_fits}
we show the hydrogen lines computed for the 
case of lower $\log{g}=2.0$. In the observed spectrum the upper members of the Bracket series are much weaker; this 
can be taken as independent evidence of the  higher \logg, as determined by M17. Indeed, our fits of the theoretical spectra 
computed for $\log{g}=4.19$ and $\log{g}=4.10$  to the observed spectra are far superior.

We also note the presence of a broad feature at 10\,700\,\AA\ (identified by ``??'' in Fig.~\ref{_fits}). 
Given the width of the feature, a possible identification is \pion{C}{i} ($^3$P$-^2$S) at 
10\,686.01, 10\,688.27, 10\,694.17\AA; these lines have
relatively high excitation potentials ($\sim7.5$~eV).

\subsection{The \pion{Ca}{ii} triplet: evidence of a chromosphere\label{DIS}}

Absorption in the  \pion{Ca}{ii} triplet (8\,498.023, 8\,542.091, 8\,662.14\,\AA) --
by comparison with other metallic lines --
is a notable feature of \kic's spectrum (see Fig.~\ref{_fits}). This Figure shows
relevant portion of Fig.\ref{_S} on a larger scale, to compare the observed \pion{Ca}{ii} lines with 
those computed for the case of a classical model atmosphere. 
The observed lines are notably weaker by comparison with the computed. 
The discrepancies between computed and observed \pion{Ca}{ii}
lines strength may be explained by several factors, such as NLTE effects or a Ca deficit.
In the former case, NLTE effects in other lines should also be apparent, but this is not the case.
Also \pion{H}{i} and metallic lines are well fitted for 
the case of solar abundances.

   \begin{figure}
   \centering
    \includegraphics[width=6.cm]{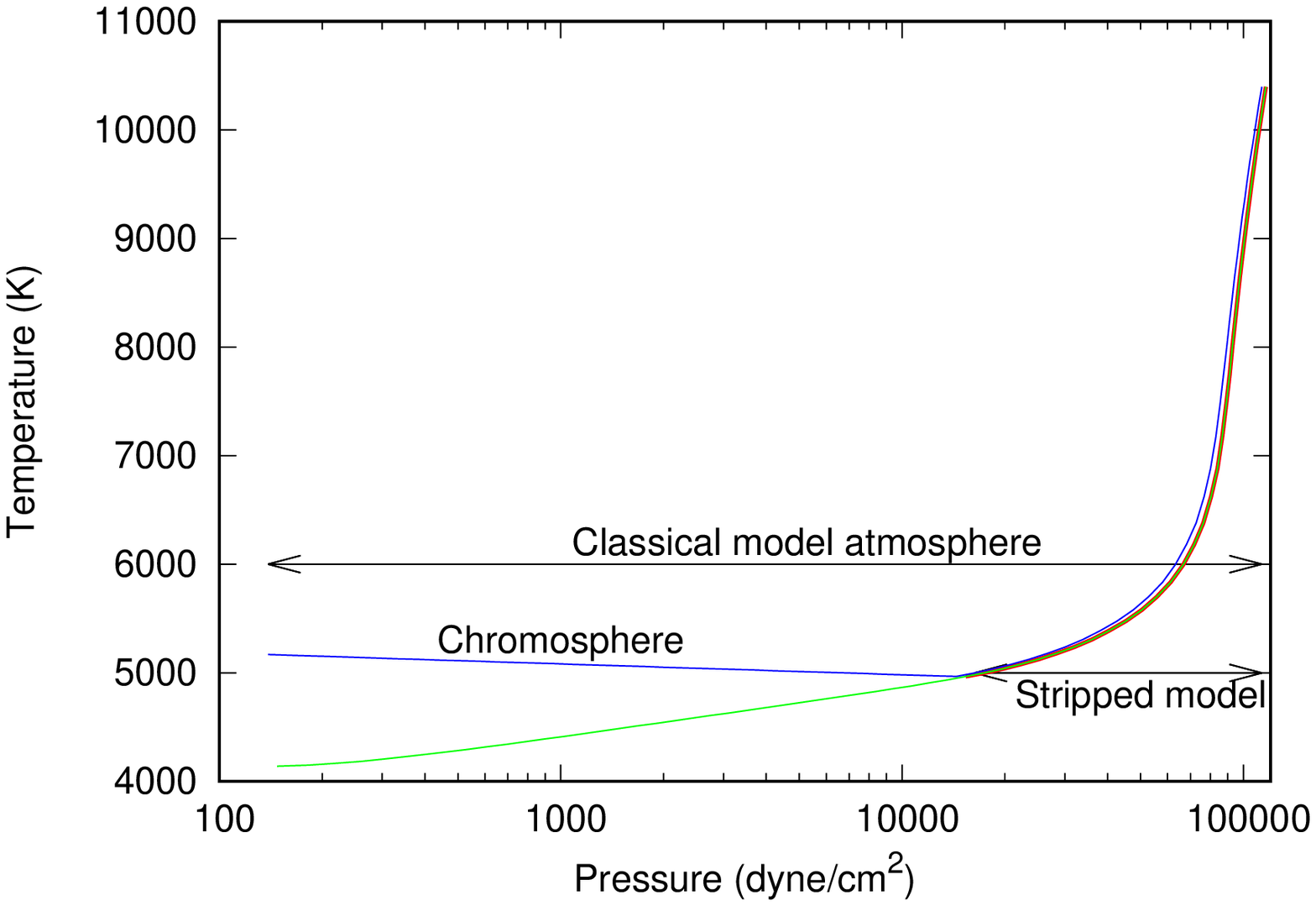}
   \includegraphics[width=6cm]{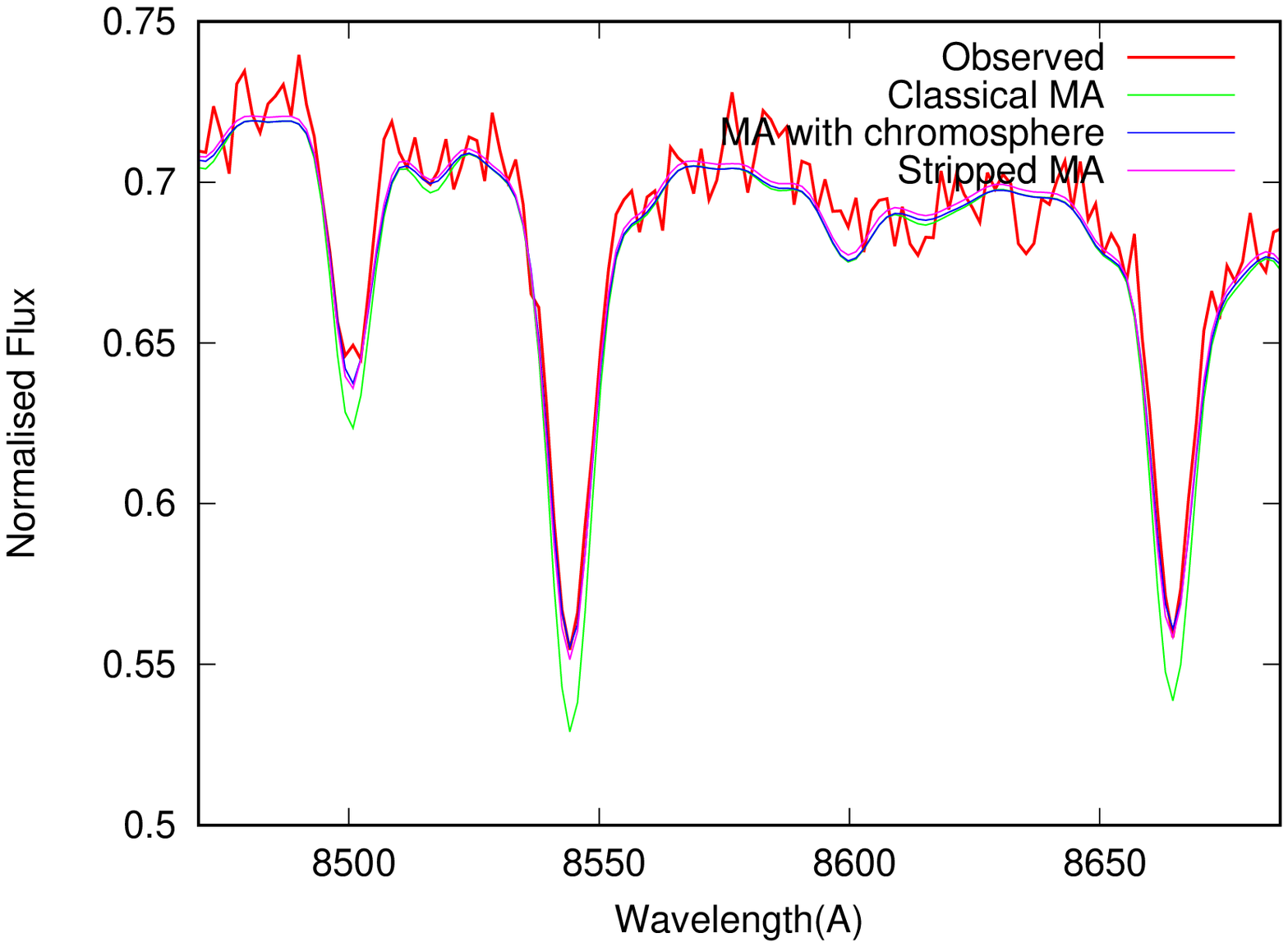}
   \includegraphics[width=6cm]{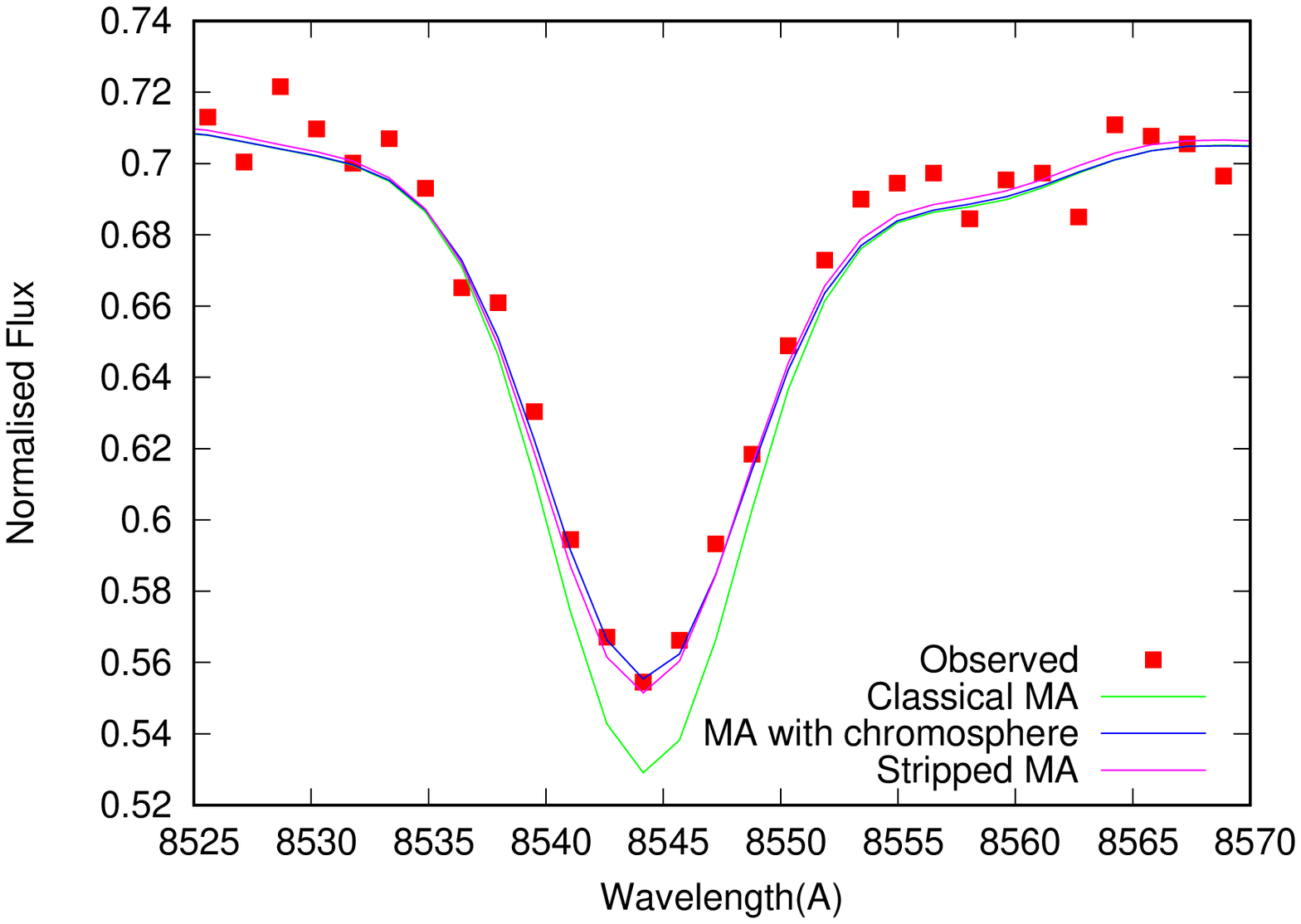}
   \caption{Top panel: the structure of the ``classical'' and ``chromospheric'' model atmosphere of component $A$.    
   Middle panel: best the fit to the observed \pion{Ca}{ii} triplet, with classical and ``chromospheric'' model atmospheres.
   Bottom panel: fit to strongest component of the \pion{Ca}{ii} triplet shown on a larger scale.
   In all three panels the ``Stripped model atmosphere'' refers to the model briefly discussed at the end of Section~\ref{DIS}.}
   \label{_caii_2}
    \end{figure}

The components of the \kic\ system consist of rapidly rotating,
tightly bound stars in a contact binary system and we would not expect the atmospheres of 
such stars so be adequtely described by the classical model. In particular
rapid stellar rotation is likely to lead to strong chromospheric (and related) activity, which
could certainly account for the weakness of the Ca triplet.

We have simulated a chromosphere by imposing a temperature minimum of
5\,000~K, and a temperature
gradient $G=dT/d(\log{m})$ ($m$ being the column density)
in the model atmosphere.
We fit the \pion{Ca}{ii} line profiles
by choosing the value of $G$ that minimises the fitting parameter $S$
\citep[details of our approach to this calculation is given in][]{pav98}.
A model with a weak chromosphere ($G=-200$, cf. the solar value $G=-800$, with $T$ in K, $m$ in 
g~cm$^{-2}$) provides 
a good fit to the data in the region of the \pion{Ca}{ii} triplet
(see Fig.~\ref{_caii_2}).

We also note here an alternative interpretion: 
our classical model atmospheres were computed for Rosseland mean optical depth in the range
$\tau_{\rm Ross} = 10^{-5}$ -- 100 but such models lead to \pion{Ca}{ii} lines that are too strong.
Interestingly, a ``stripped'' model atmosphere for the  $A$ 
component that lacks atmospheric layers in the $\tau_{\rm Ross}$ range [$10^{-5}$ -- $3\times10^{-2}$] 
provides an equally good fit to the \pion{Ca}{ii} triplet (see Fig.~\ref{_caii_2}).

However, while both approaches provide equally good fits to the data, for reasons to be discussed below
we prefer the chromospheric interpretation.


\subsection{\pion{He}{i} at 10\,830\,\AA}

A strong absorption feature is present at 10\,830\,\AA\ (see Fig.~\ref{_fits});
we attribute this to \pion{He}{i} 10\,830\,\AA\ ($^3$S$-^3$P$^0$), which is also
observed in absorption in the spectra of metal poor subdwarfs \citep{smit12} and red giants \citep{smit14}.
This line is noticeably narrower than the rotationally broadened photospheric lines, and
most likely forms outside the stellar photospheres, possibly in the common envelope.

In the case of the Sun the line is also observed in absorption; \cite{avrett} suggested that
coronal radiation penetrates into the upper solar chromosphere, causing sufficient helium 
ionisation to populate the lower level of the \pion{He}{i} 10\,830\,\AA\ transition, thus resulting in optically-thin 
absorption against the photospheric continuum. 
This, together with the presence of 
the \pion{C}{i} lines at $\sim10\,690$\,\AA, lends support to the chromospheric 
explanation offered in Section~\ref{DIS}.

\section{Conclusions}

We have carried out an analysis of the near IR spectrum of \kic. 
We computed synthetic spectra for two classical 1D model atmospheres, and fitted
the dereddened spectrum. While we acknowledge that the computed model atmospheres may be overly
simplistic for the case of the complex \kic\ system,  we do nonetheless obtain a
satisfactory reproduction of the observed SEDs. This may be taken as evidence that, at the time of observation, 
the 1D+LTE approach is valid. 

Within the constraints  of our simplified model we conclude that:

   \begin{enumerate}
\item our ephemeris analysis and spectral synthesis show that, at the time of observation,
the contribution  of the hotter component prevails in the combined spectrum;
\item the strength of the \pion{H}{i} lines is well reproduced by our models, providing
independent confirmation of the high $\log{g}$ (4.10, component $A$; 4.19, component $B$; M17) at least in the 
atmosphere of component $A$. Model atmospheres with lower $\log{g}$ result in Bracket and Paschen lines that are too srong;
\item a comparatively narrow feature at 10\,830 \AA\ is identified as the \pion{He}{i}
($^3$S$-^3$P) line, and is formed in the outer part of the common envelope;
\item a satisfactory fit to the \pion{Ca}{ii} IR triplet 
is obtained by introducing a chromosphere into the model atmosphere.
This is likely a consequence of rapid stellar rotation;
\item despite the intermediate resolution of observed spectra and the fast rotation 
      of both components, we confirm that component $A$ has near-solar abundances.
   \end{enumerate}

\begin{acknowledgements}

We warmly thank David Sand for making possible the IRTF observations, and for help with the data reduction.
The Infrared Telescope Facility is operated by the University of Hawaii under contract
NNH14CK55B with the National Aeronautics and Space administration.
We thank the referee for their helpful comments.

\end{acknowledgements}



\begin{thebibliography}{99}

\bibitem [Avrett et al.(1994)]{avrett} Avrett, E. H., Fontenla, J. M., Loeser, R., 1994, in IAU Symposium,
  154, Infrared Solar Physics, eds D. M. Rabin, J. T. Jefferies,   C. Lindsey, Kluwer Academic Publishers; Dordrecht, p.~35

\bibitem [Bond et al.(2003)]{bond03} Bond, H. E., Henden, A., Levay, Z. G., Panagia, N., Sparks, W. B., 
          Starrfield, S., Wagner, R. M., Corradi, R. L. M., Munari, U. 2003, Nature, 422, 405

\bibitem [Cushing et al.(2005)]{cushing} Cushing M. C., Rayner J. T., Vacca W. D., 2005, ApJ, 623, 1115

\bibitem [Eastman et al.(2010)]{east10} Eastman, J., Siverd, R., Gaudi, B. S., 2010, PASP, 122, 935

\bibitem[{Gray(1976)}]{gray76} Gray, D. F., 1976, The observation and analysis of stellar photospheres, Wiley
          Interscience, New York

\bibitem [Kasliwal(2012)]{kasl12} Kasliwal, M. M. 2012, PASA, 29, 482

\bibitem [Kirk et al.(2016)]{kirk} Kirk, B., et al., 2016, AJ, 151, 68

\bibitem [McCollum et al.(2014)]{mcco14} McCollum, B., Laine, S., V\"ais\"anen, P., Bruhweiler, F. C., Rottler, 
          L., Ryder, S., Wahlgren, G. M., Barway, S., Nagayama, T., Ramphul, R. 2014, AJ, 147, 11
          
\bibitem [Molnar et al.(2015)]{moln15} Molnar, L. A., Van Noord, D. M., Steenwyk, S. D., Spedden, C. J.,
          Kinemuchi, K. 2015, BAAS, 225, 415.05

\bibitem [Molnar et al.(2017)]{moln17} Molnar, L. A., Van Noord, D. M., Kinemuchi, K., Smolinski, J. P., 
         Alexander, C. E., Cook, E. M., Jang, B., Kobulnicky, H. A., Spedden, C. J., Steenwyk, S. D.
         2017, ApJ, 840, 1 (M17)

\bibitem [Nichols et al.(2013)]{nicholls} Nicholls, C. P., et al., 2013, MNRAS, 431, L33

\bibitem [Pavlenko(1998)]{pav98} Pavlenko, Ya. V. 1998, Astronomy Reports, 42, 501

\bibitem [Pavlenko(2003)]{pavl03} Pavlenko, Ya. V. 2003, Astronomy Reports, 47, 59

\bibitem [Pavlenko et al.(2008)]{pavl08} Pavlenko, Ya. V., Evans, A., Kerr, T., et al. 2008, A\&A, 485, 541

\bibitem [Pejcha(2014)]{pejc14} Pejcha, O. 2014, ApJ, 788, 22

\bibitem [Rayner et al.(2003)]{rayner} Rayner J. T., Toomey D. W., Onaka P. M., Denault A. J., 
        Vacca W. D., Cushing M. C., Rayner J. T., 2003, PASP, 115, 389
  
\bibitem [Ryabchikova et al.(2015)]{ryab15} Ryabchikova, T., Piskunov, N., Kurucz, R. L., et al. 2015, Phys. Scr, 90, 054005

\bibitem [Skrutskie et al.(2006)]{skru06} Strutskie, M. F., et al., 2006, AJ, 131, 1163

\bibitem [Smith et al.(2012)]{smit12} Smith, G. H., Dupree, A. K., Strader, J., 2012, PASP, 124, 1252

\bibitem [Smith et al.(2014)]{smit14} Smith, G. H., Dupree, A. K., Strader, J., 2014, ArXiv e-prints [arXiv:1410.3487]

\bibitem [Smith et al.(2016)]{smith16} Smith et al. 2016, MNRAS, 458, 950

\bibitem [Tylenda et al.(2011)]{tyle11} Tylenda, R., Hajduk, M., Kami\'nski, T., Udalski, A., Soszy\'nski, I., 
          Szyma\'aski, M. K., Kubiak, M., Pietrzyński, G., Poleski, R., Wyrzykowski, \L., Ulaczyk, K. 2011, A\&A, 528, A114

\bibitem [Vacca et al.(2003)]{vacca} Vacca, W. D., Cushing, M. C., Rayner, J. T., 2003, PASP, 115, 389

\bibitem [Williams et al.(2015)]{will15} Williams, S. C., Darnley, M. J., Bode, M. F., Steele, I. A. 2015, ApJ, 805, L18

\bibitem [Wright et al.(2010)]{wrig10} Wright, E. L., et al., 2010, AJ, 140, 1868

\bibitem [Zhu et al.(2013)]{zhu13} Zhu, C., L\"u, G., Wang, Z., 2013, ApJ, 777, 23


\end{thebibliography}

\end{document}